\documentclass[twocolumn,showpacs,amsmath,amssymb, prb,reprint, preprintnumbers]{revtex4-1}

\usepackage{stmaryrd}
\usepackage{txfonts}
\usepackage{amssymb}
\usepackage{mathrsfs}
\usepackage{graphicx}% Include figure files
\usepackage{dcolumn}% Align table columns on decimal point
\usepackage{bm}% bold math
\usepackage{epsfig}
\usepackage{color}                    % For creating coloured text and background
\usepackage{hyperref}                 % For creating hyperlinks in cross references

\begin{document}
\preprint{\href{http://dx.doi.org/10.1103/PhysRevB.87.184507}{S.-Z. Lin \emph{et. al.}, Phys. Rev. B {\bf 87}, 184507 (2013).}}

\title{Characterization of thin-film NbN superconductor for single-photon detection by transport measurements}

\author{Shi-Zeng Lin, Oscar Ayala-Valenzuela, Ross D. McDonald, Lev N. Bulaevskii,  Terry G. Holesinger, Filip Ronning, Nina R. Weisse-Bernstein, Todd L. Williamson, Alexander H. Mueller, Mark A. Hoffbauer, Michael W. Rabin, Matthias J. Graf}

\affiliation{Los Alamos National Laboratory, Los Alamos, New Mexico 87545, USA}

\date{\today}

\begin{abstract}
The fabrication of high-quality thin superconducting films is essential for single-photon detectors. Their device performance is crucially affected by their material parameters, thus requiring reliable and nondestructive characterization methods after the fabrication and patterning processes. Important material parameters to know are the resistivity, superconducting transition temperature, relaxation time of quasiparticles, and uniformity of patterned wires. In this work, we characterize micro-patterned thin NbN films by using transport measurements in magnetic fields. We show that from the instability of vortex motion at high currents in the flux-flow state of the $IV$ characteristic,  the inelastic life time of quasiparticles can be determined to be about $2$ ns. Additionally, from the depinning transition of vortices at low currents, as a function of magnetic field, the size distribution of grains can be extracted. This size distribution is found to be in agreement with the film morphology obtained from scanning electron microscopy and high-resolution transmission electron microscopy images.
\end{abstract}

%checked for Characterization of the NbN superconductor
 \maketitle

\section{Introduction}
The potential applications of the single-photon detector (SPD) in quantum cryptography, ultrafast photon detection experiments, and dark sky observations are very promising. In particular, superconductor-based SPDs have attracted considerable attention in the last decade.\cite{Ilin2000,Goltsman2001,Romestain2004,Kerman2006,Kitaygorsky2005,Kitaygorsky2007,Miki2008,Annunziata2009,Baek2011,Clem2011,Illin2012,Hofherr2012,Engel2012,Natarajan2012} 
Among these, research focused on the NbN superconducting nanowire SPD (SNSPD) for the following reasons: 
1) the superconducting energy gap, $\Delta$, is two orders of magnitude smaller than in semiconductor-based SPDs, which allows for the detection of low energy photons in the infrared region of the spectrum up to $5\ \rm{\mu m}$;\cite{Marsilin12} 
2) potentially fast detection times with gigahertz count rates; \cite{Ilin2000} [Currently, the reset time of devices is limited by the kinetic inductance of the superconducting nanowire and the shunt resistor, for details see Refs. \onlinecite{Kerman2006,Yang2007,Kerman09,Marsili12b}.]
3) low dark count rates are attainable, because the SNSPD is operated in a cryogenic environment; \cite{Kitaygorsky2005,Kitaygorsky2007,Bartolf2010,Yamashita2011} 4) their device efficiency is high. \cite{Natarajan2012}
So far, most of the work has  focused on the NbN SNSPD, because thin NbN has an extremely short superconducting coherence length of a few nanometers,  $\xi\sim 4$ nm, with a relatively high superconducting transition temperature, $T_c\sim 14$ K, and strong electron-phonon coupling for fast energy relaxation times. The small $\xi$ permits one to reduce the dimensionality of SNSPDs to nanoscale-sized wires for increased sensitivity to infrared photons with wavelengths $\lambda > 1.5\ \rm{\mu m}$. Since the superconducting condensation energy density per unit volume is materials specific and allows for little variability, given the constraints listed above, the obvious dimensional tunability of the device is to reduce the volume element of the detector that needs to go normal to trigger a photon count.  
Recently, the sensitivity of SNSPDs to ions at low energy and soft x-rays was explored.\cite{Sclafani2012,Inderbitzin2012}
For a review of SNSPDs based on other superconductors see e.g. Ref. \onlinecite{Natarajan2012}.

The operating principle of the NbN SNSPD is as follows.\cite{Goltsman2001,Kerman2006,Yang2007,Natarajan2012} The nanowire is biased by a DC current close to the critical current. When an incident photon interacts with the NbN nanowire, it excites a cloud of quasiparticles, that diffuses and drives a belt-like normal region across the wire. When this extended normal region appears, it expands due to Joule heating until the resistance of the NbN nanowire becomes much larger than that of a parallel shunt resistor. As a result, the current redistributes to the shunt and a voltage pulse is detected. The process of heat diffusion and transition of belt-like region to the normal state is very fast and takes place within $\sim 10$ ps for a 100 nm wide wire. It takes much longer for the normal region in the nanowire to recover back to the superconducting state and for the bias current to flow back into the nanowire. The redistribution of the current at this stage is slow (1-10 ns) due to the large kinetic inductance of the NbN nanowire and the shunt resistance. \cite{Kerman2006,Yang2007,Kerman09,Marsili12b} During this time the nanowire cools down to the bath temperature (with phonon escape time $\sim 160$ ps \cite{Semenov95}) and the SNSPD is again ready for the detection of incoming photons. In the absence of incident photons, some part of the nanowire may become normal as well because of thermal fluctuations, which cause the detection of so-called dark counts.\cite{Engel2006,Kitaygorsky2007,Bartolf2010} 
It was proposed that the dominant contribution to the dark count rate is from the crossing of single vortices in the NbN nanowire due to thermal fluctuations in the metastable DC-biased superconducting state.\cite{Bulaevskii2011,Bulaevskii2012}     

The operation of the NbN SNSPD involves the excitation and relaxation of quasiparticles, a complex nonequilibrium problem. The excited quasiparticles relax into the equilibrium state through electron-electron scattering, electro-phonon scattering and recombination of quasiparticles into Cooper pairs. \cite{Kaplan1976} The relaxation of quasiparticles is characterized by a relaxation time $\tau$, which  plays an important role in determining the physically limiting SNSPD performance. On the other hand, grain boundaries are inevitably introduced during the growth process of thin NbN films. These boundaries work as pinning centers for vortices, and thus may affect the vortex crossing in nanowires and their dark count rate. We expect that knowledge of the inelastic relaxation time of quasiparticles and the size distribution of grains in thin films are important for the device optimization of NbN SNSPDs.   

In this work, we measure the standard materials properties given by the normal-state resistivity $\rho(T)$ and superconducting transition temperature $T_c$. These are supplemented by the extraction of the inelastic relaxation time $\tau$ of quasiparticles and the distribution of grain sizes $P(L)$ in thin NbN superconductors from transport measurements of the $IV$ characteristics. The former is extracted from the instability of vortex motion in the flux-flow state, while the latter is obtained from the depinning transition of vortices. The inelastic relaxation time of quasiparticles in the vortex state is found to be about $\tau\sim 2$ ns in our NbN films. 
The nondestructive determination of the grain size distribution is dominated by domains of linear dimension of less than $5 \xi$ with an exponential tail for domains larger than $\sim 50 \xi$. This result is confirmed by morphology studies of the NbN film with electron microscopy.

The remaining part of the paper is organized as follows. Sec.~II describes thin film growth, fabrication, micropatterning, morphology analysis, and standard transport characterization. In Sec.~III.A, we discuss the instability of the flux-flow state and inelastic quasiparticle relaxation time. This is followed by Sec.~III.B, with the investigation of  the depinning transition and nondestructive extraction of grain size distribution. The paper concludes with a short summary in Sec.~IV.

\section{Thin film growth and characterization}
 
\begin{figure}[t]
\psfig{figure=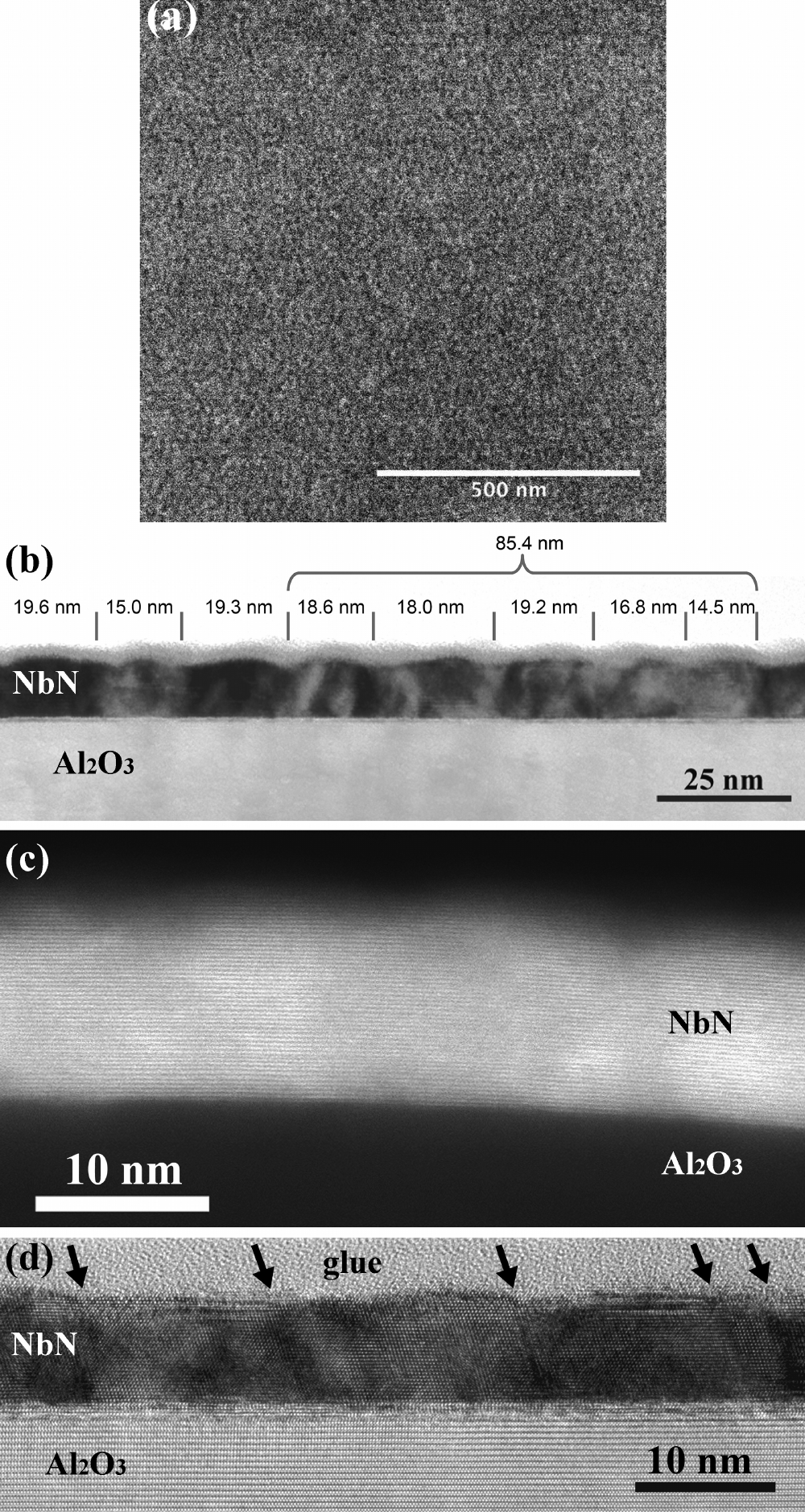,width=1\columnwidth}
\caption{
The morphology of ENABLE-grown NbN film is visualized in 
(a) with the $1 \times 1$ $\mu{\rm m}^2$ plane-view SEM image. The film thickness of $d=11.2\pm 0.4$ nm and 
grain size and film morphology are shown in (b) with the bright-field TEM cross-sectional image.
Bending of the lattice fringes around surface irregularities is shown by the Z-contrast STEM imaging in (c).
The TEM image of a sister sample in (d) shows the atomic structure of the interface and grain boundaries in these NbN films.
}
\label{fig:fig1}
\end{figure} 

\begin{figure}[t]
\psfig{figure=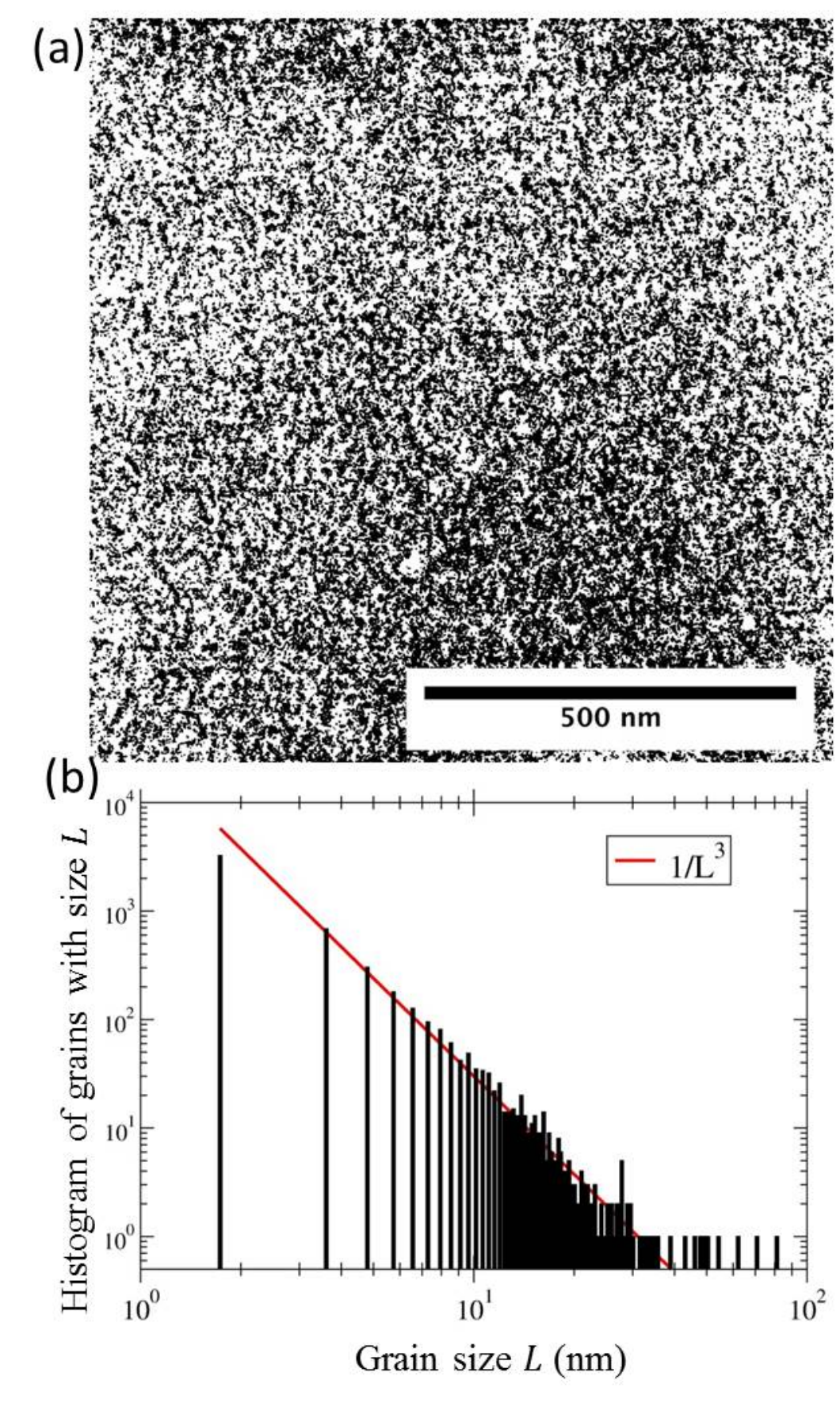,width=0.9\columnwidth}
\caption{(color online).
Grain-size analysis of the SEM image shown in Fig.~\ref{fig:fig1}(a).
Panel (a): Post-processed SEM image where gray scales have been mapped onto a binary black and white map. Black domains are delineated by white "background" for follow-on analysis.
Panel (b): The ``Analyze Particles'' method in {\em ImageJ} \cite{imageJ,imageJdoc,Dunes2011} was used to create the
histogram of the grain-size distribution of the black domains in the $1\times 1 \, \mu {\rm m}^2$ area shown in (a). For small domains the distribution obeys a $\sim 1/L^3$ behavior (red solid line) with minimal linear dimension of  $L_{min} = 1.75$ nm. %The grains cover 47\% of the area.
} 
\label{fig:fig2}
\end{figure}

%%
% Newly added part by Terry
%%

\subsection{Growth and fabrication}

The thin NbN films in this study were grown by a molecular-beam-epitaxy-type growth process called Energetic Neutral Atom Lithography and Epitaxy (ENABLE). ENABLE utilizes an energetic beam of neutral N atoms (kinetic energies of 1 to 5 eV) to activate nitride thin-film growth. The high energy and reactivity of N atoms allow for growth of high-quality, uniform crystalline thin films with high yield that are difficult to grow by conventional chemical vapor deposition (CVD), molecular beam epitaxy (MBE) techniques or magnetron sputtering. ENABLE has previously been used for the growth of III-nitride semiconductors.\cite{Williamson11b,Williamson11,Mueller06}
The thin film was grown on a 50 mm diameter substrate of c-axis sapphire wafer that was first pre-nitrided at 400$^\circ$ C under the ENABLE N-atom beam for 20 minutes. Following pre-nitridation, the NbN film was grown at 600$^\circ$ C by having the Nb metal flux and N atom beams concurrently bombard the substrate. The Nb flux was provided by a Nb rod in a miniature electron-beam cell manufactured by Mantis Deposition Ltd (Oxfordshire, UK). The resulting NbN film was cooled to ambient temperature in vacuum and was $d=11.2$ nm thick. The face-centered cubic crystallographic structure of NbN was confirmed with X-ray diffraction. Finally, the fabricated NbN micron-sized wires were photolithographically defined using MicroChemicals AZ 5214E photoresist and Microposit MF~319 developer. The pattern defined in the photoresist was transferred into the underlying film using a 50\% chlorine in argon inductively coupled plasma (ICP) etch. The contact pads were patterned via a lift-off technique with the metal (5 nm Ti, 200 nm Au) deposited by an electron-beam evaporator. The wafer was then diced using a resin bonded diamond blade. 

\subsection{Film structure and morphology}

Structural characterization of the films was accomplished with scanning electron microscopy (SEM) and transmission and scanning transmission electron microscopy (TEM and STEM). SEM and TEM were carried out in either a FEI Tecnai F30 Twin S/TEM or FEI Titan 80-300 S/TEM, both operated at 300 kV.  The film morphology of ENABLE-grown films is shown in Fig.~\ref{fig:fig1}. The SEM image in panel (a) shows the uniformity of the film in a $1 \mu{\rm m}\times 1\mu{\rm m}$ view. The contrast giving rise to the grainy morphology arises from the 1 to 2 unit-cell deep grooves above the grain boundaries in the film. This grain boundary morphology is shown in the TEM cross-sectional view of this film in Fig.~\ref{fig:fig1}(b).   
The average thickness of the film from the TEM measurements was $d=11.2\pm 0.4$ nm.  The image was taken near a zone axis.  Some of the NbN grains are crystallographically aligned to the beam to create highly-diffracting conditions and thus appear dark in the bright-field TEM images. Nearby grains misoriented by only a few degrees are not in the highly diffracting conditions and appear light in contrasts.  Of particular note is the grain boundary grooving observed between these grains.  The grain boundary morphology or grooving is used to define grain sizes in the other parts of the image where adjacent grains are not distinguishable by diffraction contrast.  From this contrast and microstructure, we infer grain sizes in this image to be on the order of 15-20 nm.  
The Z-contrast STEM image of the Fig.~\ref{fig:fig1}(c) shows the lattice fringes of the NbN film and the ability of the film to conform to surface irregularities on the substrate.  The high-resolution TEM image of Fig.~\ref{fig:fig1}(d) was taken from a sister sample to the sample shown in Fig.~\ref{fig:fig1}(a)-(c).  In this image, it appears that the top layers of the substrate have been modified by the pre-nitridation step used in the film preparation.  Laterally, the film interface is abrupt and well-defined.   Grain boundaries can again be identified by changes in the atomic structure and the grain boundary grooving described above.  Grain sizes in this image are in the 5-20 nm range and are typical of the films grown in this series. The grain boundary structures shown here are important for providing the collective pinning of the vortex lattice in the superconducting state and determine the value of the critical current, as will be revealed in the next section in the analysis of the $IV$ characteristics. 

The grooving along grain boundaries can provide sufficient contrast in high magnification, plane-view SEM images to enable a much broader analysis of grain sizes in these films.  An in-depth analysis of the large 1 $\mu{\rm m}^2$  SEM image in Fig.~\ref{fig:fig1}(a) is performed after converting the gray-scale image of different grain orientations to a binary black and white image in Fig.~\ref{fig:fig2}(a). This allows a better delineation of grains (black domains) for further identification. Because the identification of domains depends on the specific threshold used for creating a black and white image one should keep in mind that our quantitative analysis provides only an approximate count of domains that in principle should be verified by a series of TEM cross-sectional images.
For a quantitative analysis of domains in the large 1 $\mu{\rm m}^2$ view area, we used the ``Analyze Particles'' method of the image tool {\em ImageJ} \cite{imageJ,imageJdoc,Dunes2011} to generate a histogram of grain size area $A$ of the black domains with bins of size 10 nm$^2$. The corresponding linear size, $L=\sqrt{A}$, histogram is shown in Fig.~\ref{fig:fig2}(b). $L$ varies over several orders of magnitude between 1.75 nm and 81.2 nm with a mean value of $L_{\rm{avg}} = 6.7$ nm.
The grains that were mapped onto black domains in Fig.~\ref{fig:fig2}(a) account for roughly 47\% of the completely covered film. 

The key result of the quantitative SEM analysis is that the distribution of grains is dominated by small grains, which are reasonably well described by an inverse power-law $1/L^3$, as shown by the red solid line in Fig.~\ref{fig:fig2}(b). 
Notably the majority of grains extracted from the large view SEM image is less than 20 nm, which is consistent with the TEM cross-sections shown in Fig.~\ref{fig:fig1}(b) and (d). 
The importance of the peculiar inverse power-law will reappear in the analysis of the depinning current of the flux lattice state in Sec.~III.B.

%\clearpage

%%%

\subsection{Resistivity}

\begin{figure}[t]
\psfig{figure=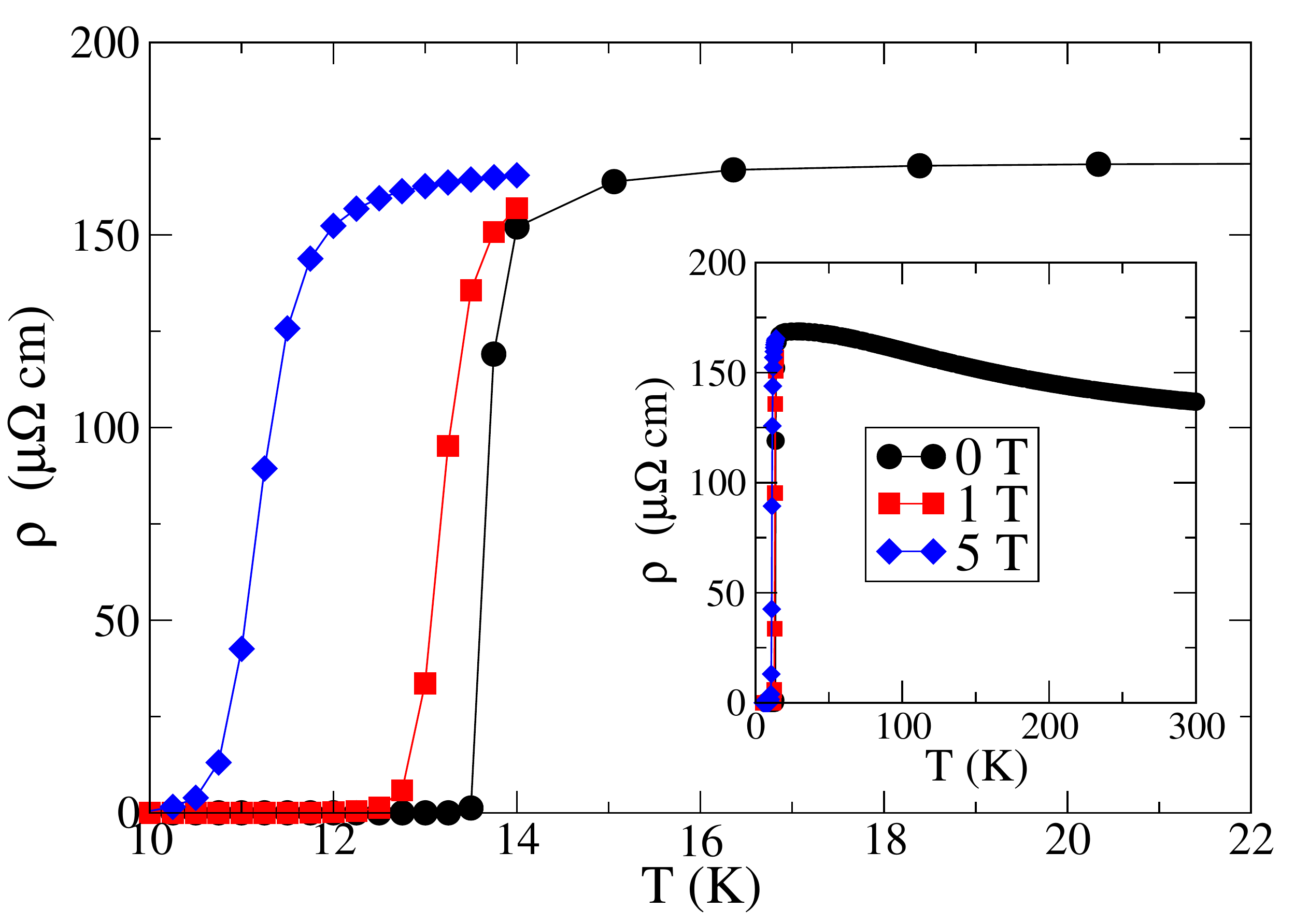,width=1.0\columnwidth} 
\caption{(color online). Resistivity $\rho$ vs.\ temperature $T$ in magnetic fields $B=0, 1, 5$ T. 
Inset:  $\rho(T)$ between room temperature and $T_c$ is characteristic of a \emph{bad metal}
with a RRR value less than one, ${\rm RRR} =\rho({\rm 300\, K})/\rho(T_c)=0.83$.
}
\label{fig:resistivity}
\end{figure}

Standard film characterization was performed by transport measurements in the micron-sized wire using four-point probe technique in Physical Property Measurement System (PPMS).
Figure \ref{fig:resistivity} shows the resistivity data close to the superconducting transition $T_c=13.7\pm 0.2$ K at zero-applied magnetic field $B=0$ T.  $T_c$ was determined by the midpoint of the resistive transition, with error bars determined by the temperatures at which the resistance was at the $10\%$ and $90\%$ value of the normal state.
The resistivity is in agreement with reports for thicker films in Ref.~\onlinecite{Chockalingam2008}.
The transition is suppressed with the field applied perpendicular to the film. At $B=1$ T and 5 T we observe $T_c(B) = 13.2$ K and $11.2$ K, respectively. From this field dependence we estimate for the slope of the upper critical field  at the phase transition, $d B_{c2}/dT=-2.0$ T/K, and derive a zero-temperature coherence length 
$\xi(0)= \sqrt{-\Phi_0/(2\pi T_c d B_{c2}/d T)}= 3.5$ nm, where $\Phi_0$ is the flux quantum.
In addition, in superconductors where the mean-free-path $\ell$ of electrons is shorter than the zero-temperature coherence length $\xi(0)$, i.e.,  superconductors in the {\it dirty} limit,  the diffusion constant of electrons can be obtained directly from
$D=4 k_B/(\pi e \, d B_{c2}/d T) = 0.55\ \rm{cm^2/s}$, 
where $k_B$ is the Boltzmann constant and $e$ is the negative electron charge.

We use electronic structure calculations of the electronic dispersion along high-symmetry directions to estimate the Fermi velocity in bulk NbN 
to be of the order $v_f \approx 100 - 150$ km/s.\cite{Fong1972,Klintenberg2012,Klintenberg,Ortiz2009}  
Then from the diffusion coefficient
$D=v_f \ell/3 =  0.55\ \rm{cm^2/s}$, we
obtain for the mean-free-path $\ell \approx 1.0 - 1.5 \ \rm{nm}< \xi(0)$,
which is roughly three to ten times larger than reports for disordered thick films ($d>50$ nm) grown by magnetron sputtering.\cite{Chand2009} 
We can perform a consistency check to see whether these values compare reasonably well with a rough estimation from superconducting parameters by 
employing $v_f \sim \pi\xi_0 \Delta/\hbar$.
If we assume $v_f \approx 150$ km/s, the superconducting gap $\Delta \approx 3$ meV,\cite{Hajenius2004,Chockalingam2009,Kamlapure2010}   
and the relation between the clean and dirty limit  coherence length in superconductors \cite{deGennes}
$\xi_0 \simeq 1.4\, \xi(0)^2/\ell$, then for $\xi_0 = 3\xi(0)$
the mean-free-path is $\ell\approx 0.46\, \xi(0)=1.6$ nm. 
In conclusion, all these estimates are in agreement with each other and our earlier analysis of the film morphology of ENABLE grown NbN films, which points toward a superconductor in the dirty limit. 

The inset in Fig.~\ref{fig:resistivity} shows the temperature dependence of the resistivity $\rho(T)$ up to 300 K for a wire of size $l \, (\text{length})\times w \, (\text{width})\times d \, (\text{thickness})=100\rm{\ \mu m}\times 5\ \mu m\times 11.2\ nm$. We extract a resistivity ratio of RRR=0.83 between room temperature and slightly above $T_c$, which is indicative of charge transport in a \emph{bad metal}, where defect and grain boundary scattering are important.
Such a scenario is consistent with a short mean-free-path $\ell \alt \xi_0$ and superconductivity in the dirty regime.

NbN thin films with different thickness grown on sapphire were investigated systematically in Ref. \onlinecite{Semenov2009}. For films with similar thickness, the $T_c$ of our samples is about $1.5$ K lower, while other quantities such as diffusion constant $D$, electron mean-free-path $\ell$, and zero-field superconducting critical current $j_d(0)$ (see Sec. III B) are similar to those reported in  Ref. \onlinecite{Semenov2009}.

\section{Transport measurements in the flux-flow state}

\begin{figure}[t]
\psfig{figure=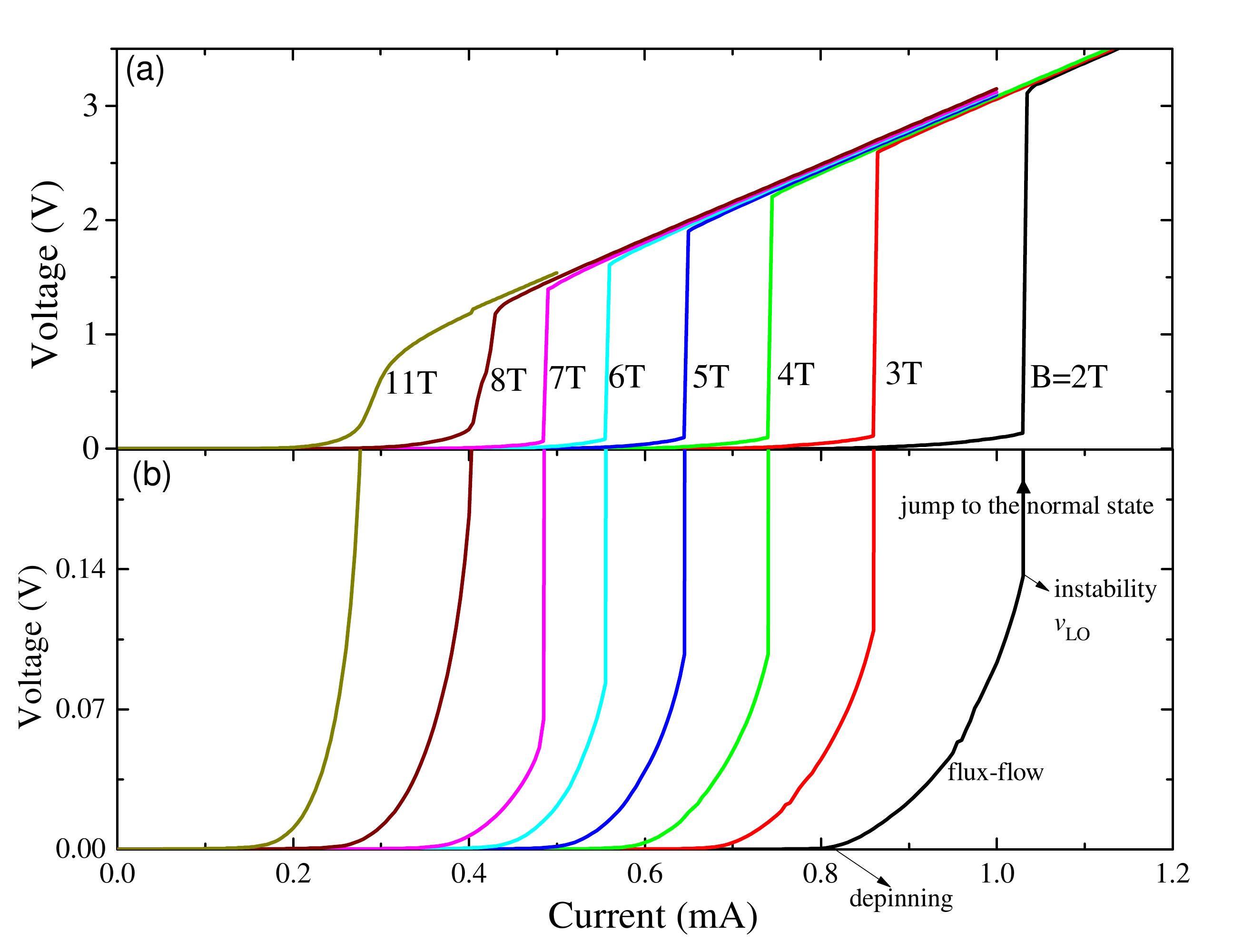,width=\columnwidth} 
\caption{\label{f1} (color online). (a): \emph{IV} curves at several representative magnetic fields. Three different regions in the \emph{IV} characteristics can be clearly seen in (b). Below the depinning current $I_d$, voltage vanishes and the lines overlap with the $x$-axis.}
\end{figure} 

In type-II superconductors ($\lambda\gg\xi$) vortices enter the superconductor above the lower critical field $B_{c1}$. Under an applied DC transport current, vortices are driven by the Lorentz force perpendicular to the current, which is balanced by the pinning force in inhomogeneous superconductors as long as the transport current is smaller than the depinning current. In this case, vortices do not order into a vortex lattice state due to pinning at inhomogeneities. Finally, when the transport current exceeds the depinning current, vortices  move, resulting in the flux-flow state. In the flux-flow state, the inhomogeneities in the superconductor are quickly averaged out by vortex motion and the lattice order is recovered.\cite{Koshelev94,Besseling03} 

In our experiments, sample environments at fixed temperature down to 2.3 K were controlled in a Janis flow cryostat in low DC magnetic fields, from 0 to 0.1 T, generated by an electromagnet at room temperature and in a single shot fridge in an Oxford superconducting magnet with a Variable Temperature Insert for fields between 0.1 T and 15 T. The critical current measurements were carried out by means of two different pulsed techniques to avoid damage, self-heating and/or thermal runaway of the samples. We varied pulse duration and duty cycle as a mean of assessing and minimizing the self-heating. In the first method the commercial set of Keithley instruments, Nanovoltmeter Model 2128A and Model 6221 AC/DC Current Source, was used in synchrony to create the characteristic \emph{IV} curves of the superconducting NbN films at fixed magnetic fields and temperatures. In the second method two waveform generators were used to create a periodic signal applied to the sample through a shunt resistor used to record the current. Voltage was measured from the corresponding sample leads, both signals were linearly amplified and stored in a scope. Both pulsed techniques show results in excellent agreement. 
The dimensions of the NbN film are $l \, (\text{length})\times w \, (\text{width})\times d \, (\text{thickness})=100\rm{\ \mu m}\times 5\ \mu m\times 11.2\ nm$. All  $IV$ transport measurements were performed at $T=2.3$ K. For our analysis we use  $\xi(0)\approx 3.5\rm{\ nm}$. 
In addition, from the London penetration depth, $\lambda(0)\approx 410$ nm, reported for 
films of similar thickness,\cite{Kamlapure2010} we deduce the Pearl length $\Lambda(0)\equiv 2\lambda^2/d\approx 30\rm{\ \mu m}$ relevant for screening of magnetic flux in thin films.
Typical \emph{IV} curves at several magnetic fields are depicted in Fig.~\ref{f1}. Three different regions can be clearly seen. Below the depinning current $I_d$ vortices are pinned and the superconductor is in the zero-voltage state. Above the depinning current, the vortex lattice moves in the flux-flow state causing dissipation. At a critical current $I_{\rm{LO}}$ (voltage $V_{\rm{LO}}$) an instability occurs and the superconductor switches to the normal state. The dependence of $I_d$ and $I_{\rm{LO}}$ on the external magnetic field is nonlinear, and both decrease with field as $-\ln(B)$ as will be discussed below.
As $B\rightarrow 0$, $I_d$ and $I_{\rm{LO}}$ become the same and they are close to the depairing current.

\subsection{Larkin-Ovchinnikov instability}

The jump at $I_{\rm{LO}}$ is due to the instability of the collective motion of the vortex lattice as predicted by Larkin and Ovchinnikov (LO) several decades ago.\cite{Larkin1975}  The instability is related to the quasiparticle relaxation, thus one can extract the inelastic quasiparticle relaxation time in the magnetic field from the instability. 
The argument for this effect is as follows: as vortices move, an electric field is induced in the normal core of vortices resulting in dissipation. Additionally, the electric field shifts the distribution of quasiparticles and pushes them
outside the normal core. As a consequence, the size of the vortex core shrinks
\begin{equation}\label{eq1}
\xi^2(v)=\frac{\xi^2(0)}{1+v^2/v_{\rm{LO}}^2}.
\end{equation}  
where $\xi^2(0)$ is the coherence length at the velocity of vortex $v=0$ and $v_{\rm{LO}}$ is the critical velocity of the instability. Because of the reduction of the  size of the vortex core $\xi(v)$ at a velocity $v$, the Bardeen-Stephen viscosity also decreases
\begin{equation}\label{eq2}
\eta^2(v)=\frac{\eta^2(0)}{1+v^2/v_{\rm{LO}}^2}.
\end{equation} 
For a given Lorentz force, the increase of vortex velocity diminishes $\eta(v)$, and hence increases $v$ even further. This positive feedback speeds up the vortex motion and renders the flux-flow state unstable at a critical velocity
\begin{equation}\label{eq3}
v_{\rm{LO}}^2=1.31\frac{D }{\tau} \sqrt{ 1-T/T_c}.
\end{equation} 
where $D=v_f \ell/3$ is the quasiparticle (electron) diffusion constant with Fermi velocity $v_f$. Thus one can determine $\tau$ by measuring the instability velocity of the flux-flow state. The LO instability of the flux-flow state has been observed both in conventional \cite{Klein1985,Samoilov1995,Ruck1997, Ruck2000, Armenio2007,Cirillo2011} and high-$T_c$ cuprate superconductors.\cite{Doettinger1994,Doettinger1997}

\begin{figure}[t]
\psfig{figure=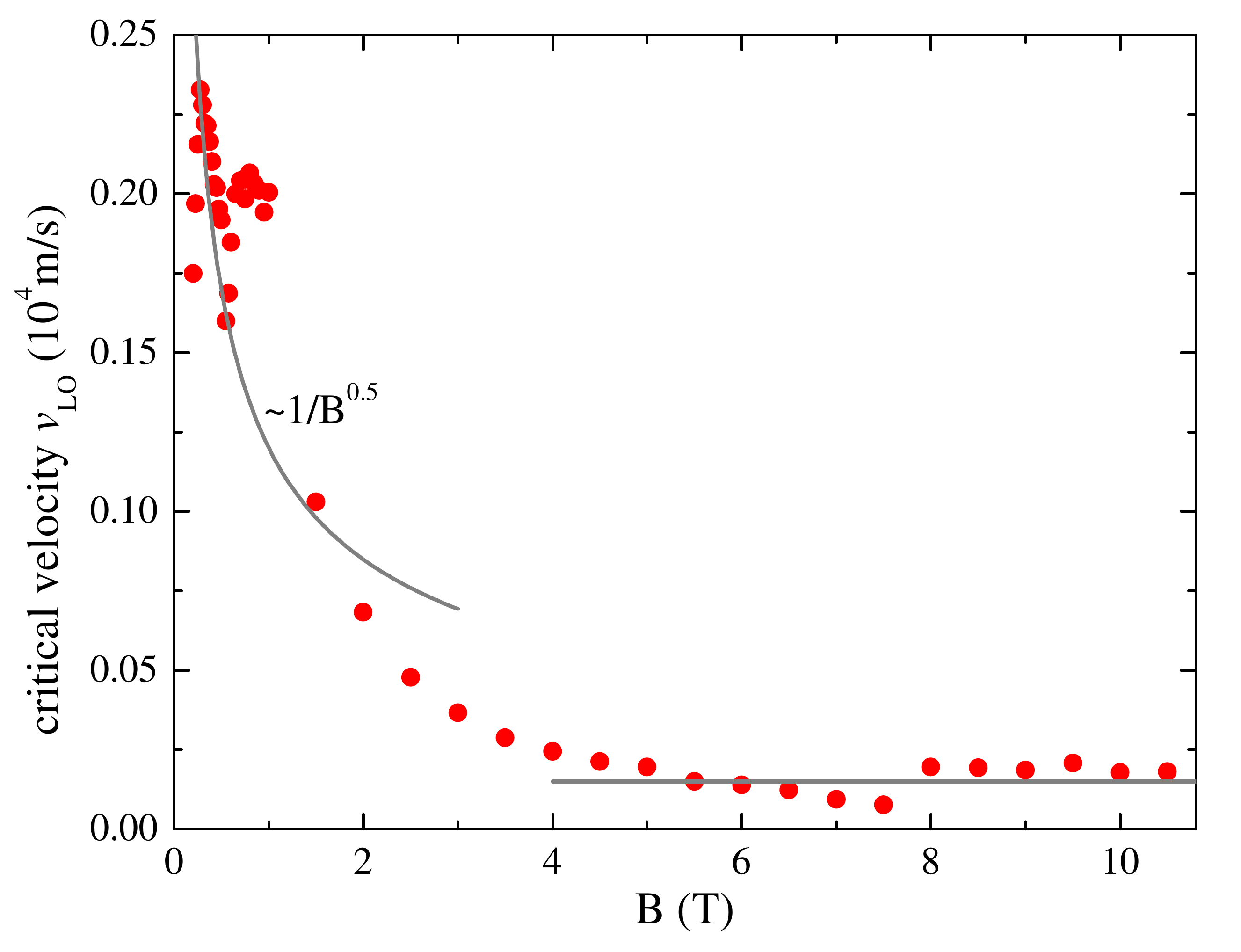,width=\columnwidth} 
\caption{\label{f2} (color online). 
Dependence of the critical velocity on magnetic field in the LO instability region. In the low-field region $v_{\rm{LO}}\sim 1/\sqrt{B}$, whereas  in the high-field region $v_{\rm{LO}}$ is independent of $B$. The solid lines are guide to the eyes.}
\end{figure}

Typical \emph{IV} curves at several magnetic fields are depicted in Fig. \ref{f1}. Above the depinning current $I_d$, vortices move giving rise to the flux-flow state. The resistance in the flux-flow state increases with current because of the shrinkage of the vortex core as a result of the nonequilibrium LO effect. At a critical current (voltage), the system switches to the normal state. The critical velocity $v_{\rm{LO}}$ is given by $v_{\rm{LO}}=V_{\rm{LO}}/(\mu_0 B l)$ with $V_{\rm{LO}}$ the voltage at the end of the flux-flow branch, $B$ the applied magnetic field, and $\mu_0$ the vacuum permeability. At higher field, the transition to the normal state becomes smooth due to the increase of vortex viscosity. We extract $V_{\rm{LO}}$ from the $IV$ characteristics using the criteria that at $V_{\rm{LO}}$ the derivative $d V/dI$ jumps.

\begin{figure}[t]
\psfig{figure=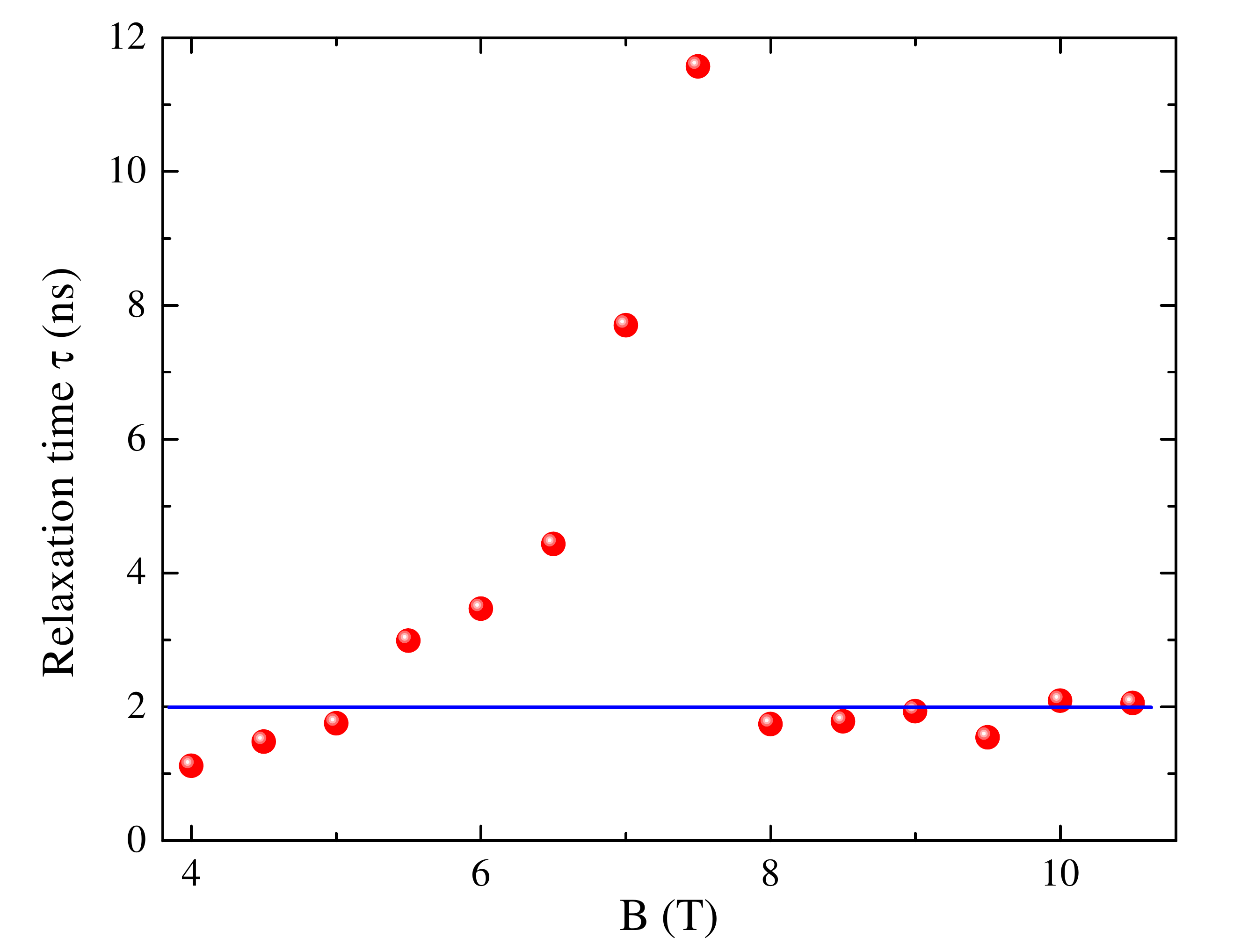,width=\columnwidth} 
\caption{(color online). Extracted inelastic quasiparticle relaxation time $\tau$ is independent of $B$ in the high field region.}
\label{fig:tau}
\end{figure} 

The dependence of $v_{\rm{LO}}$ on $B$ is shown in Fig. \ref{f2}. It decreases as $1/\sqrt{B}$ for a weak $B$ and then saturates at a constant value. By assuming a uniform distribution of quasiparticles in superconductors, LO predicted that the critical velocity $v_{\rm{LO}}$ is independent on the applied magnetic field, see Eq. (\ref{eq3}). The uniform distribution of quasiparticles is realized at high magnetic fields where the inter-vortex distance $a$ is small, such that $v_{\rm{LO}} \tau \gg a$ with $a\approx \sqrt{\Phi_0/{B}}$. In the low magnetic field region, Eq. (\ref{eq3}) becomes inapplicable, because the quasiparticle distribution at $v_{\rm{LO}}^*$ given by Eq. (\ref{eq3}) is nonuniform and is confined inside the unit cell of the vortex lattice, i.e. $v_{\rm{LO}}^* \tau\ll a$. When the velocity of vortices increases such that quasiparticles are no longer confined in the unit cell of vortex lattice, i.e. the condition $v \tau \gg a$ is fulfilled, the flux-flow instability is triggered \cite{Doettinger1995}. Therefore $v_{\rm{LO}}\sim a/\tau$ in the low magnetic field region, as shown in Fig. \ref{f2}.  A magnetic-field independent critical velocity $v_{\rm{LO}}$ in the high-field region indicates that the heating effect due to vortex motion is weak and can be neglected. In the opposite case of large self-heating, it was found theoretically\cite{Bezuglyj1992} and experimentally \cite{Kunchur2002,Peroz2005} that $v_{\rm{LO}}$ decreases as the magnetic field increases, $v_{\rm{LO}}\sim 1/\sqrt{B}$, which is clearly not the case here.

In the next step, we use Eq.~(\ref{eq3}) to find the inelastic relaxation time for quasiparticles at high fields, as shown in Fig.~\ref{fig:tau}. 
The plateau of $\tau\approx 2\ \rm{ns}$ above 8 T is expected for flux-flow dominated by the LO instability at high fields, while the origin of the rise in $\tau$ between 6 T and 8 T is not understood at this time.
A similar plateau in the relaxation time of NbN was reported in Ref.~\onlinecite{Cirillo2011}. Obviously, for technical applications a shorter relaxation time for excited quasiparticles is preferred in order to achieve a faster response of the NbN SNSPD after the formation of a hot spot. For that reason the authors of Ref.~\onlinecite{Cirillo2011} fabricated NbN/ferromagnetic hybrids, where $\tau$ is two orders of magnitude smaller than in conventional NbN film due to the additional scattering channel of quasiparticles by magnetic impurities in the ferromagnetic layer, leading to a faster relaxation of excited nonequilibrium quasiparticles.

The quasiparticle relaxation time at high magnetic fields can be extracted from the LO instability of flux flow. It might be interesting to ask how the relaxation time at high fields extrapolates to the zero field case, where usually the SNSPD is operated. Recently, time-resolved, optical pump-probe measurements on a thin $\rm{Nb_{0.5}Ti_{0.5}N}$ film show that the quasiparticle relaxation time (around 1 ns) depends weakly on the magnetic fields up to $8$ T. \cite{Zhang2006} Thus one expects that the quasiparticle relaxation time at zero magnetic field has the same order of magnitude as the one extracted from the LO instability at high fields. 

\begin{figure}[t]
\psfig{figure=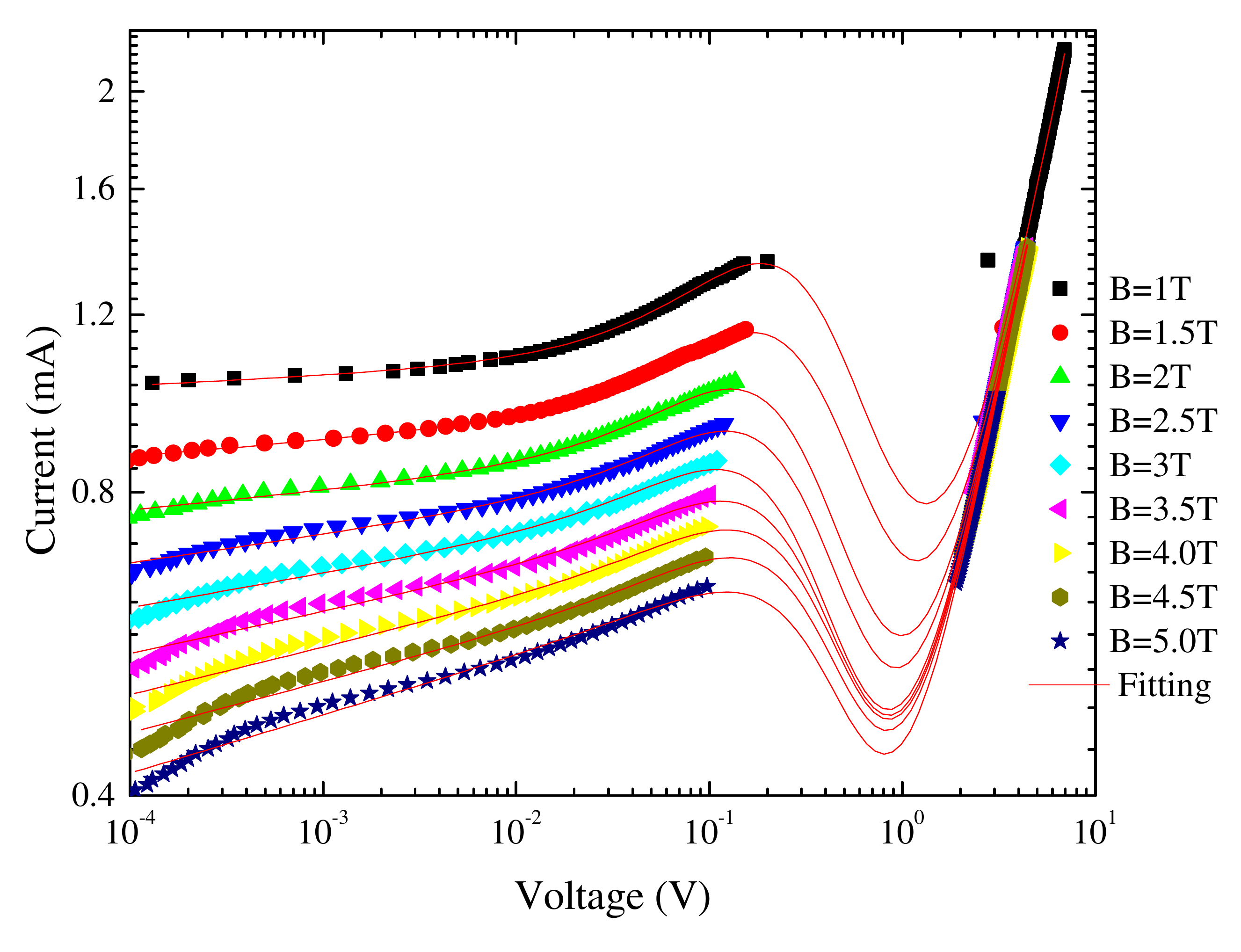,width=\columnwidth} 
\caption{\label{f4} (color online). LO instability fits (lines) according to Eq.~(\ref{eq4}) of experimental (symbols) \emph{VI} curves. The gap in the plot means that the transition from the flux-flow state to the normal state is very sharp.}
\end{figure} 
\begin{figure}[b]
\psfig{figure=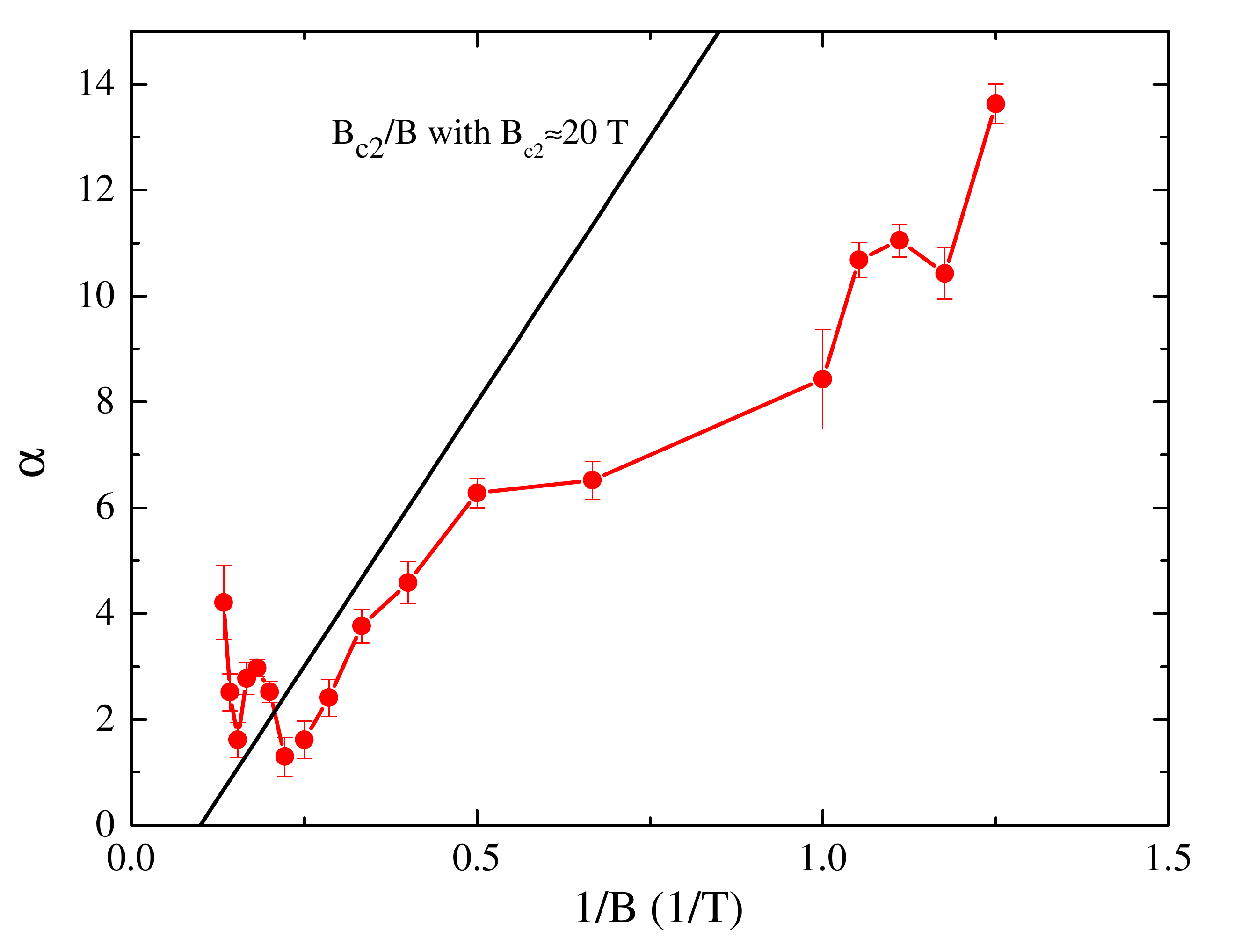,width=\columnwidth} \caption{\label{fa4} (color online). Dependence of $\alpha(B)$ on $B$ obtained from fits of Eq.~(\ref{eq4}) to experimental data.
The solid line is a guide-to-the eye for the Bardeen-Stephen flux-flow behavior for $B_{c1} \ll B < B_{c2}$. The upper critical field $B_{c2}(0)=-0.69 T_c \frac{d B_{c2}}{d T}|_{T_c}\approx 20$ T is estimate using the Werthamer-Helfand-Hohenberg expression \cite{Werthamer1966} with the transport measurements from Fig. \ref{fig:resistivity}. }
\end{figure}

The \emph{IV} curves including the LO instability can be described by the following phenomenological equation \cite{Ruck2000}
\begin{equation}\label{eq4}
I(V)=\frac{V}{R_n}\left[\frac{\alpha(B)}{1+V^2/V_{\rm{LO}}^2}+\frac{\beta(B) (V/V_{\rm{LO}})^{-c}}{1+V^2/V_{\rm{LO}}^2}+1\right] ,
\end{equation}
where the first term in the bracket accounts for the reduced Bardeen-Stephen viscosity in the nonequilibrium region, the second term accounts for the pinning effect, and the last term is the damping due to the suppression of superconductivity around the vortex core. Here $R_n\approx 2.8\ \rm{k\Omega}$ is the normal-state resistance at $T_c$ and $\alpha$, $\beta$, $c$ are fit parameters that depend on $B$. In the limit $V\ll V_{\rm{LO}}$, we should recover the linear $IV$ curve for $I-I_d=V/R_f$ where $R_f=R_n B/B_{c2}$ is the Bardeen-Stephen flux-flow resistance. By expanding Eq. \eqref{eq4} with respect to $V/V_{\rm{LO}}\ll 1$ and comparing with the linear $IV$ curve, we thus obtain $c=1$ and $\alpha(B)=B_{c2}/B$. 
Our experimental data can be fitted very well by Eq.~(\ref{eq4}) as presented in Fig.~\ref{f4}. The fit parameter $\alpha(B)$ is shown in Fig.~\ref{fa4}, where $\alpha(B)= B_{c2}/B$ as expected for high fields not too close to $B_{c2}$ and from the limit of $V/V_{\rm{LO}} \ll 1$. The other fit parameter is $c\approx 1$. From the fitted curves, an unstable branch of the \emph{IV} curve near the LO instability becomes visible and the system develops hysteresis around the instability region. The hysteretic \emph{IV} curve, due to the LO instability, has also been observed experimentally in Ref.~\onlinecite{Samoilov1995}.

\subsection{Vortex pinning at  grain boundaries}

\begin{figure}[t]
\psfig{figure=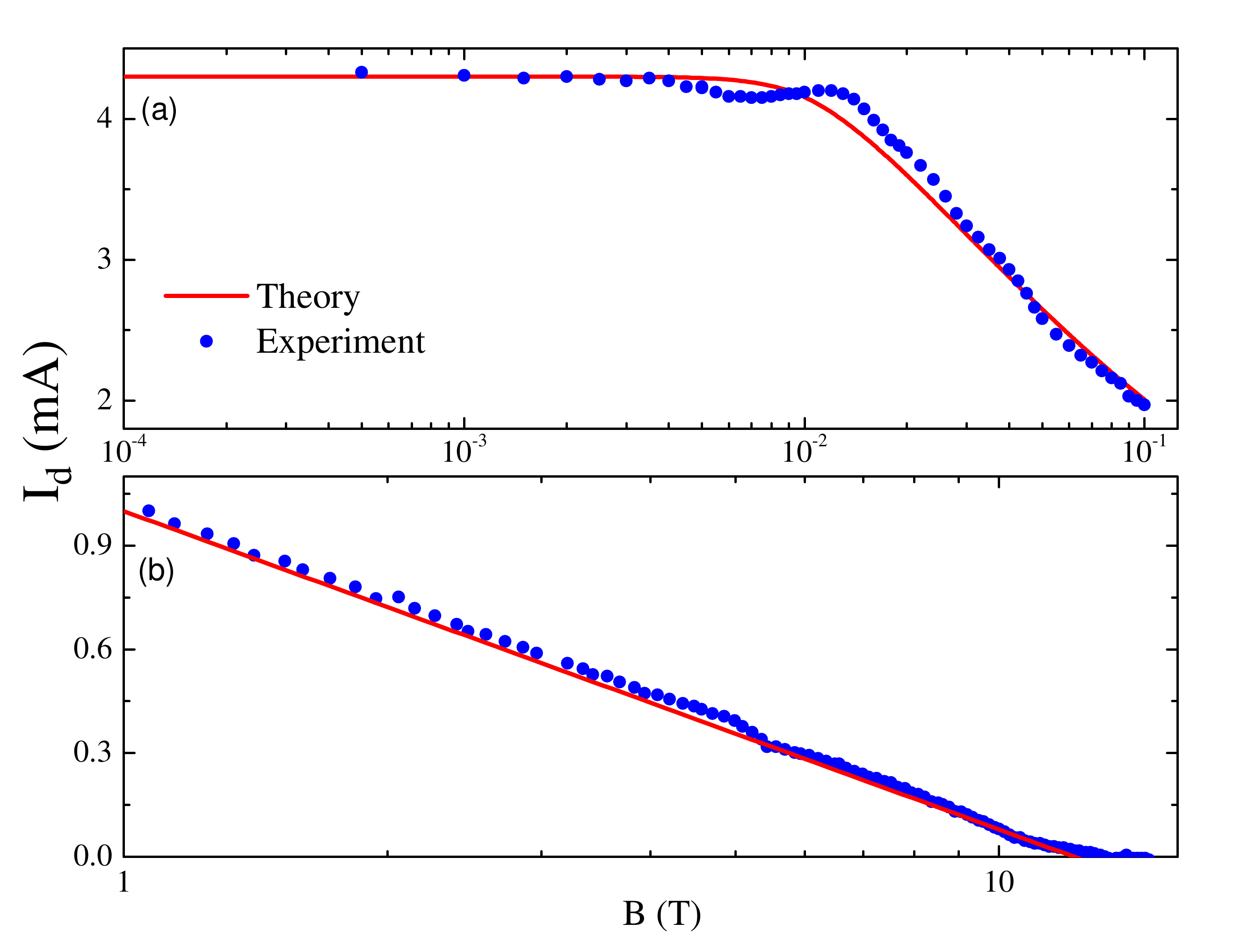,width=\columnwidth} 
\caption{\label{f5} (color online). Experimental depinning current $I_d$ as a function of $B$ (symbols) at low DC magnetic fields (a) and high magnetic fields (b). The high field data are measured in a single-shot fridge in an Oxford superconducting magnet.  Lines are theoretical curves obtained by using Eq. (\ref{eq:Id}) and the distribution function in Fig. \ref{f6}.}
\end{figure}

\begin{figure}[b]
\psfig{figure=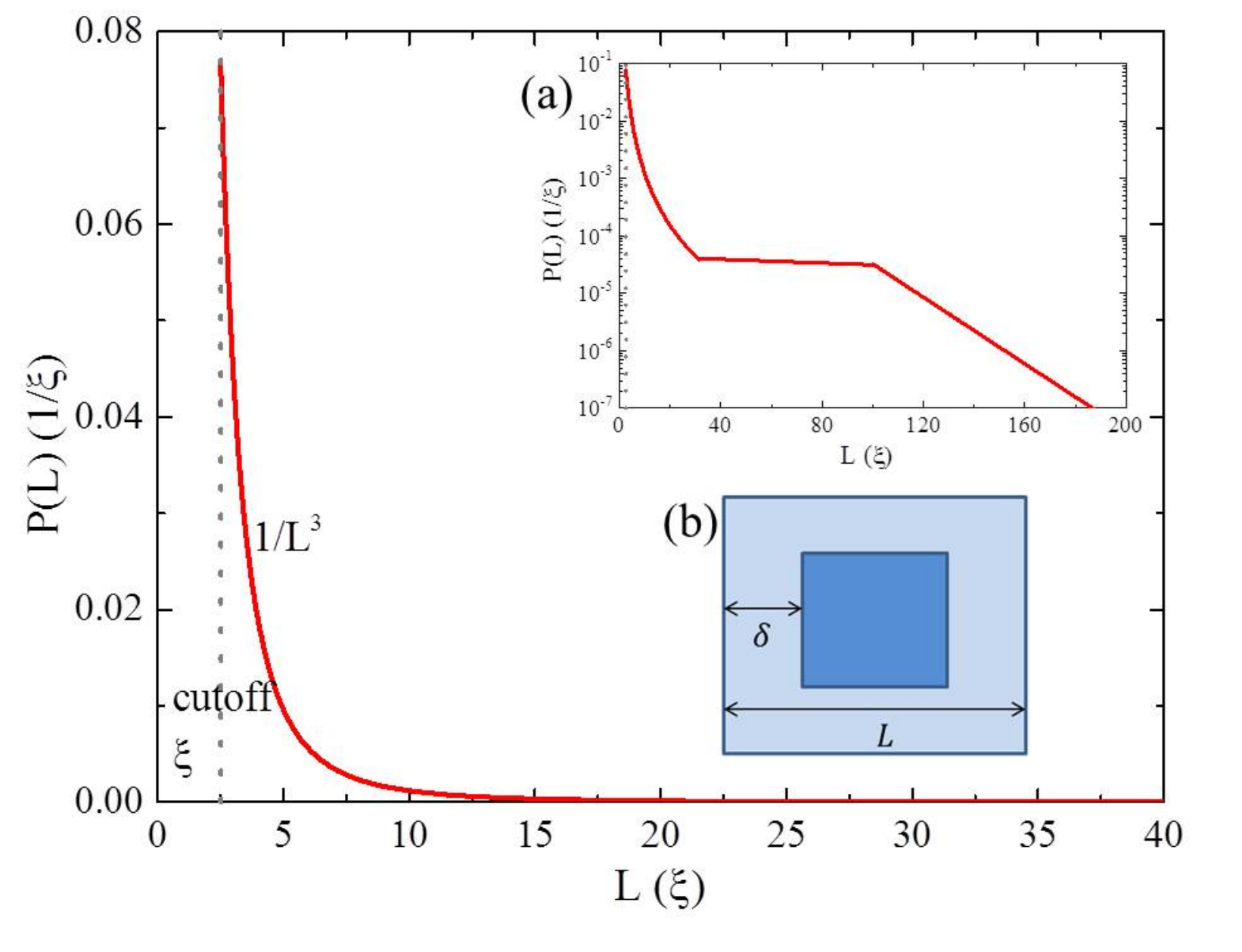,width=\columnwidth} 
\caption{(color online). Size distribution of grains obtained from Eq. (\ref{eq:dis}). Inset (a): Semi-logarithmic plot of the same distribution. Inset (b): Schematic view of a square grain with size $L$ and pinning area of width $\delta$.}
\label{f6}
\end{figure} 

We proceed with the investigation of the depinning transition of vortices. We use the practical criterion that a depinning transition occurs when the measured voltage $V$ is larger than an arbitrary threshold of 1 mV. From that we can obtain the dependence of the depinning current $I_d$ on magnetic field from the \emph{IV} curves. The results are compiled in 
Fig.~\ref{f5}. The depinning current $I_d$ depends weakly on $B$ when $B<10$ mT and decreases as $I_d\sim -\ln B$ above. It is worth to note that this logarithmic dependence cannot be explained by the collective pinning theory,\cite{Blatter94} which predicts $I_d\sim B^{-2}$ for thin films. 

It is known from thin-film growth parameters and confirmed by TEM and SEM images that our NbN films exhibit island-like growth resulting in granular morphology. Since superconductivity is suppressed at grain boundaries, they provide a pinning potential for vortices and thus may affect SNSPD detector performance. The pinning due to grain boundaries has also  been observed experimentally in high-temperature superconductor $\rm{YBa_2Cu_3O_{7-\delta}}$ films,\cite{Fedotov2002} where the dependence of $I_d$ on $B$ is similar to that in Fig.~\ref{f5}.  The depinning current density at $B=0$ T of our thin film is $j_d(0)=I_d(0) / (w d) = 7.7$ MA/cm$^2$, which is close to the depairing current density $j_c(0)=c \Phi_0/(12\pi^2 \sqrt{3} \lambda^2\xi(0))\approx 16\ \rm{MA/cm^2}$. Thus the pinning of vortices by grain boundaries can achieve a depinning current that is close to the depairing current.

The depinning current depends crucially on the size distribution of superconducting grains (domains), which can be extracted from the dependence of $I_d$ on $B$. We use a model for collective breakaway of pinned vortices developed by Fedotov \emph{et al.}\cite{Fedotov2002} 
For simplicity we further assume square domains in our thin-film NbN. A pinning theory for more general shapes of domains was presented in Ref.~\onlinecite{Pan2006} and the results are qualitatively similar to those with square domains. 
One starts with a given probability density distribution $P(L)$ of grains with linear dimension $L$, where the probability density for a vortex inside the square domain of size 
$L$ is $W(L) = \mathcal{N} L^2 P(L)$ with normalization constant $\mathcal{N}$.  The probability for finding a single vortex pinned in a domain is normalized to unity, 
$\int_0^\infty W(L) dL = 1$.

Not all vortices can sit at the energy minimum of the pinning potential due to the competition between the pining energy and elastic energy of vortex lattice. The resulting vortex configuration is a compromise between these two energies. 
Assuming a square pinning potential with strength $\epsilon_p={\Phi_0^2 r_c^2}/{(8\pi\lambda\xi_0)^2}$ and using the expression for the elastic energy of vortex lattice $\epsilon_e=\Phi_0 B \delta^2/(8\pi\lambda)^2$, we obtain the maximal displacement for pinned vortex, 
$\delta(B)=\sqrt{4\pi  r_c^2 B_{\text{c2}}/{B}}$, i.e., if the displacement of the vortex from the grain boundary is less than $\delta$, then the vortex remains pinned. Here $r_c\sim\xi_0$ characterizes the strength of pinning potential and $\delta$ is deviation from the perfect lattice. Under these conditions the vortex core gains condensation energy of superconductivity over the elastic deformation energy
by staying at the grain boundary.
It follows that the probability of a vortex to lie less than a distance $\pm\delta$ away from the grain boundary is approximately equal to the ratio of the area of four strips of width $\xi$ to the total area $L^2$ [see the inset (b) in Fig. \ref{f6}]
\begin{align}
{P}(L; \delta)=\left\{
\begin{array}{cl}
 1, & {\rm{if\ }}\ L\leq 2\delta , \\
1-\frac{(L-2\delta)^2}{L^2}, & {\rm{if\ }}\ L>2\delta .
\end{array}
\right.
\end{align}
For this case it was shown that the depinning current, normalized to its value at $B=0$, is simply the ratio of pinned vortices $n_p$ to the total number of vortices $n_{tot}$,
\begin{eqnarray}\label{eq:Id}
\frac{I_d(B)}{I_d(0)}  &=& \frac{n_p}{n_{tot}} = \int_0^\infty W(L) {P}(L; \delta)  dL
\nonumber \\
&=& 1-  \mathcal{N} \int_{2\delta }^{\infty } P(L) (L-2\delta )^2 dL . 
\end{eqnarray}
Here the magnetic field enters only through the vortex displacement $\delta(B)$.

Since a weak magnetic field corresponds to soft elastic shear stiffness of the vortex lattice, i.e., $\delta \sim \sqrt{\Phi_0/4B}$, low-field measurements probe primarily large domains, i.e. $L\gg \sqrt{\Phi_0/4B}$. From the measured $I_d$ as a function $B$, we can obtain the distribution function from Eq. (\ref{eq:Id}) by taking the third derivative with respect to $y=2\delta$ to attain
\begin{equation}\label{eq:dis}
P(y)= \frac{1}{2  \mathcal{N} I_d(0)}\frac{d^3 I_d(y)}{dy^3}.
\end{equation}
The resulting $P(y)$ is just the distribution function of the grain sizes $y$. To get rid of the small oscillations in Fig. \ref{f5}, we first smooth the experimental data and then use  Eq.~(\ref{eq:dis}). The smoothed curves (not shown) are very close to the lines in Fig. \ref{f5}. The resulting distribution function is shown in Fig. \ref{f6}, which is in reasonable agreement with the experimental one shown in Fig. \ref{fig:fig2}. The probability distribution in Fig. \ref{f6} for small grains is  much larger than that for larger grains. The distribution function for small grains follows $1/L^3$, while for large grains, it follows an exponential distribution. Since the smallest length scale for vortex is of order of $\xi$, we cannot resolve the distribution for grain sizes smaller than $\xi$ from the measurements in the flux-flow state. Finally we confirm the agreement between the measured and calculated critical current by inserting the extracted distribution of Fig. \ref{f6} back into Eq. (\ref{eq:Id}), as shown in Fig. \ref{f5}.

\section{Summary}

In summary, we characterized the quality and uniformity of ENABLE-grown thin-film NbN superconductors for potential SNSPD applications.  From transport measurements we derived superconducting material parameters $T_c=13.7$ K, $\xi(0)=3.5$ nm, and depinning current density $j_d(0)=7.7$ MA/cm$^2$. In addition, we determined that our thin films of thickness $d=11.2$ nm are in the dirty limit with the mean-free-path much shorter than the coherence length of the hypothetically clean superconductor, $\ell\approx 0.15 \xi_0$. This length scale was further corroborated by the distribution of grain sizes extracted from the analysis of SEM and TEM images and the field-dependence of depinning currents. The presented vortex theory successfully explained the collective vortex lattice motion in the flux-flow state with the Larkin-Ovchinnikov instability in the $IV$ characteristics at high bias currents, as well as the depinning current $I_d$ at low bias currents. The detailed analysis of the Larkin-Ovchinnikov instability revealed a relatively long inelastic quasiparticle relaxation lifetime of order $\sim 2$ ns, which might provide the bottleneck for the hot spot relaxation in NbN-based SNSPD devices. Finally, from the nondestructive measurement of the depinning current with magnetic fields, we extracted the characteristic domain size distribution of grains, which resulted in comparable values to the independent, yet destructive, analysis using TEM techniques. While the prevalence of grain boundaries in thin-film NbN superconductors crucially affects transport properties like critical currents, their potential for vortex pinning at low bias currents is negligible for SNSPD applications, which are typically biased close to the critical current. This study has shown the potential use of field-dependent measurements of the depinning current in micron-sized wires for determining the grain size distribution in thin-film superconductors. The advantage of a nondestructive characterization method of the uniformity of thin superconducting films may prove beneficial for the pre-screening of films for further nano-patterning.

\section{Acknowledgement}
This work was performed under the auspices of the U.S. DOE contract No. DE-AC52-06NA25396 through the LDRD program at Los Alamos National Laboratory, the Center for Integrated Nanotechnologies, an Office of Science User Facility operated for the U.S. DOE, and the National High Magnetic Field Laboratory, which is jointly supported by the U.S. DOE, NSF, and the State of Florida.

%\bibliography{reference}

\end{document}